\documentclass[
reprint,
superscriptaddress,
longbibliography,
amsmath, 
amssymb,
aps,
prb
]{revtex4-2}

\usepackage{graphicx} 
\usepackage{bm} 
\usepackage{dcolumn}
\usepackage{physics} 
\usepackage{color} 

\usepackage[colorlinks=true, allcolors=blue]{hyperref} 

\begin{document}
\setcounter{MaxMatrixCols}{12} 

\title{Dirac semimetal phases in chiral carbon nanoscrolls}

\author{Tzu-Ching Hsu}
\affiliation{Department of Physics, National Taiwan Normal University, Taipei, Taiwan}

\author{Jhih-Shih You}
\affiliation{Department of Physics, National Taiwan Normal University, Taipei, Taiwan}

\author{Hsiu-Chuan Hsu}
\affiliation{Graduate Institute of Applied Physics, National Chengchi University, Taipei, Taiwan}

\author{Ion Cosma Fulga}
\affiliation{Leibniz Institute for Solid State and Materials Research Dresden, Helmholtzstr. 20, 01069 Dresden, Germany
}
\affiliation{W{\"u}rzburg-Dresden Cluster of Excellence ct.qmat}


\begin{abstract}
Chirality induced by rolling a two-dimensional material into a spiral geometry reshapes its electronic band structure. 
In this work, we theoretically investigate the topological properties of carbon nanoscrolls under an axial magnetic field, focusing on structures in which chirality is encoded through shifted edge alignments.
In contrast to unshifted structures, where mirror symmetry pins the Dirac cones to half a flux quantum, chiral carbon nanoscrolls lack this symmetry, and Dirac cones emerge at magnetic flux values away from half a flux quantum.
We demonstrate that these Dirac cones are topologically protected by combined inversion-time reversal symmetry and remain robust even when sublattice symmetry is broken. 
Furthermore, we show that the number of Dirac cones and their real-space probability distributions depend on the number of turns and the magnetic field strength. 
Our study elucidates the role of chirality in the band topology of nanoscroll geometries.
\end{abstract}

\maketitle

\section{Introduction} 
\label{sec:introduction}

Carbon nanostructures have attracted tremendous attention for decades due to their unique electronic properties and high stability~\cite{Ayanda2024-og, Yang2015-ri, RevModPhys.81.109}.
In graphene, Dirac cones in the band structure are associated with massless relativistic-like charge carriers exhibiting high mobility and low resistivity~\cite{RevModPhys.81.109, Novoselov2005-do, Zhang2005-yu, Novoselov2007-mc}. 
These characteristics make graphene a promising platform for electronic devices such as Hall-effect sensors and field-effect transistors~\cite{10.1007/978-3-319-05353-0_60, Schwierz2010-lp}.
The ability to control Dirac cones in carbon nanostructures is therefore crucial for tailoring their electronic response and potential applications.

Since Dirac cones in graphene originate from its honeycomb lattice structure~\cite{PhysRev.71.622}, structural modifications within the graphene family can alter their number and locations in the Brillouin zone.
Additional Dirac cones can emerge in monolayer graphene when third-nearest-neighbor hopping is considered~\cite{PhysRevB.83.115404}. 
In Bernal bilayer graphene, interlayer coupling can decompose a single Dirac cone into four while conserving the topological charges~\cite{Seiler2024-py, PhysRevB.75.155424}.
In metallic carbon nanotubes, band crossings can be viewed as one-dimensional projections of graphene Dirac cones. Depending on the chiral vector, these crossings may vanish in the electronic structure~\cite{RevModPhys.79.677, Jorio2008-gq, AVOURIS2002429}.
By cutting a carbon nanotube along its axis and placing the end atoms on top of each other, one constructs a one-round carbon nanoscroll (see Fig.~\ref{fig:scrolls3d}), which exhibits a semiconducting phase with a small bandgap. However, under an axial magnetic field, band crossings appear, leading to enhanced conductance~\cite{Zhong2025-td}.

Due to the unique geometry of carbon nanoscrolls, they show potential in a variety of applications, such as supercapacitors, nanoactuators, and nanolight sources~\cite{Liu2018-zu, PhysRevB.74.085414, Han2025-fo}.
Since the idea of carbon nanoscrolls was first proposed in 1959~\cite{Bacon1960-xy}, various fabrication methods have been developed to produce high-quality carbon nanoscrolls~\cite{Carotenuto2013-ri, Huang2011-lv, Sharifi2013-wu, Xie2009-lh, Li2018-ky}.
Recent experimental studies have successfully realized carbon nanoscrolls with different roll-up directions, leading to chiro-electronic properties~\cite{Zhang2025-vd}. 
A recent theoretical work proposed a model for large-radius carbon nanoscrolls by introducing a modified version of periodic boundary conditions at opposite edges of a bilayer graphene ribbon~\cite{Zhong2022-rt}. 
Within the tight-binding approximation, it was shown that, in the presence of an axial magnetic flux equal to half a flux quantum, the band structure of the AB-stacked model with zigzag edges and mirror symmetry exhibits band crossings.
These crossings are topologically protected by a nonzero winding number~\cite{Lee2025-gm}.

In this work, within the large-radius (small-curvature) approximation, we construct a tight-binding model for chiral carbon nanoscrolls as realized through various edge-alignment configurations. 
Numerical calculations in this study are performed using Kwant~\cite{Groth2014-ii}.
We find that the band crossing no longer appears at half of the magnetic flux quantum due to the mirror symmetry breaking. 
By treating the momentum and magnetic flux as parameters, we construct a fictitious Brillouin zone (FBZ) to characterize the emergence of Dirac cones in the band structure.
The empirical relation between the momentum and magnetic flux of the Dirac cones is identified numerically and derived analytically. 
The wave functions of the cones are then analyzed.
In addition, the Dirac cones are shown to persist even when sublattice symmetry is broken by next-nearest-neighbor hopping.
Finally, we demonstrate that the combination of inversion and time-reversal~($IT$) symmetry is responsible for the topological protection of these Dirac cones in chiral carbon nanoscrolls.

The rest of this work is organized as follows.
We begin by detailing the tight-binding geometry in Sec.~\ref{sec:models}.
Afterwards, we describe the Dirac points occurring first in one-round~(Sec.~\ref{sec:1round}), and then in multi-round~(Sec.~\ref{sec:multi_round}) nanoscrolls, focusing on the magnetic fluxes at which they appear as well as on their topological protection.
We conclude in Sec.~\ref{sec:conclusions}.

\section{Tight-binding geometry} 
\label{sec:models}

\begin{figure}[tb]
\centering
\includegraphics[width=\columnwidth]{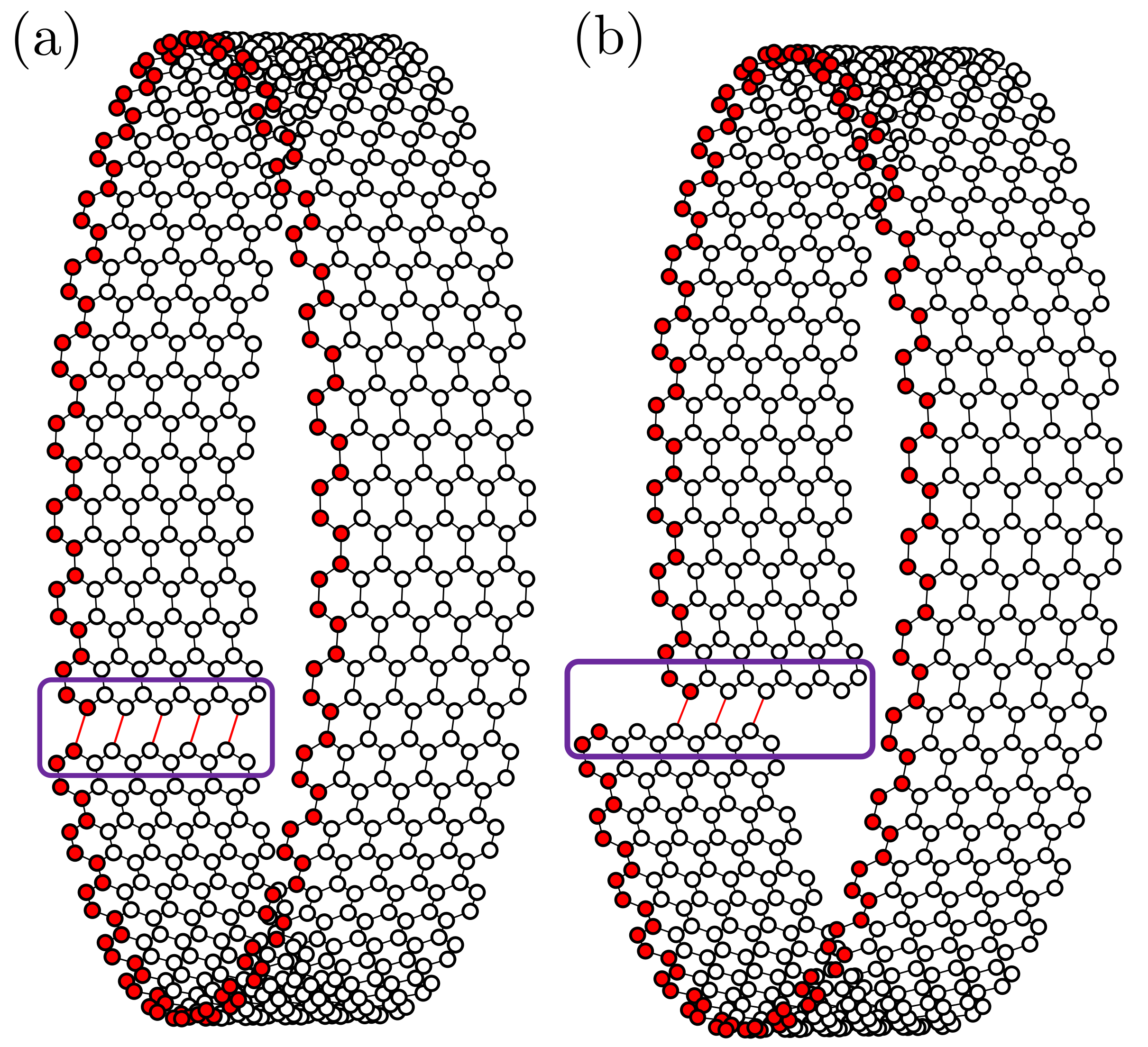}
\caption{
Sketch of two one-round nanoscrolls. The black lines indicate intralayer, nearest-neighbor hopping $r_0$, whereas the red lines indicate the interlayer hopping $r_1$. 
The purple rectangle highlights the difference in boundary alignment: no shift in panel (a), with a shift in panel (b).
In each panel, the red sites correspond to one unit cell, which is repeated infinitely many times along the axial direction in order to produce an infinitely long nanoscroll.} 
\label{fig:scrolls3d}
\end{figure}

Carbon nanoscrolls possess a unique geometric structure. 
In this work, we focus on the horizontal shift along the axial direction, which causes different boundary alignments at the edges. 
The schematic diagrams of one-round carbon nanoscrolls with zigzag edges but different boundary alignments are shown in Fig.~\ref{fig:scrolls3d}.

The theoretical results of the unshifted structures have been reported in Refs.~\cite{Zhong2022-rt, Zhong2025-td, Lee2025-gm}. 
Similar to these works, we focus on the structure with zigzag edges, adopt the large-radius (small-curvature) assumption, and model the graphene as a set of spinless orbitals on a honeycomb lattice with nearest-neighbor hopping. 
Due to the large-radius approximation, all intralayer hoppings are equal, as are all interlayer hoppings.
Thus, the carbon nanoscrolls can be modeled as multilayer, spinless graphene ribbons within a tight-binding framework subject to the mixed boundary condition~\cite{Zhong2022-rt}. 
In this work, we employ ABC stacking for the multilayer structure, since Bernal stacking is incompatible with the mixed boundary condition, and ABC stacking is energetically favored over AA stacking~\cite{Aoki2007-mj}.
We take the value of the nearest-neighbor intralayer hopping as $r_0=3.16$ eV, and the nearest interlayer hopping as $r_1=-0.381$ eV~\cite{PhysRevB.80.165406}. 
The negative sign of $r_1$ reflects the hybridization between different parts of the $p_z$ orbitals compared to $r_0$.

Carbon nanoscrolls are structurally similar to carbon nanotubes, particularly in the one-round configuration. 
The key distinction arises from the edge-to-edge hopping, which is interlayer in carbon nanoscrolls and intralayer in carbon nanotubes. 
The horizontal shift along the axial direction shown in Fig.~\ref{fig:scrolls3d} reconfigures the two atoms previously linked by interlayer hopping (the red lines in the purple rectangles) to form new interlayer connections with other atoms.
This situation is analogous to the chiral carbon nanotubes, so this geometry is termed a chiral carbon nanoscroll.
We therefore adopt the chiral vector concept from carbon nanotubes to characterize these structures~\cite{saito1998physical}. On a flattened graphene sheet, the two sites that can be connected by a chiral vector will align on top of each other once the structure is rolled up. The chiral vector in our chiral carbon nanoscroll system is defined as follows:

\begin{figure}[tb]
\centering
\includegraphics[width=\columnwidth]{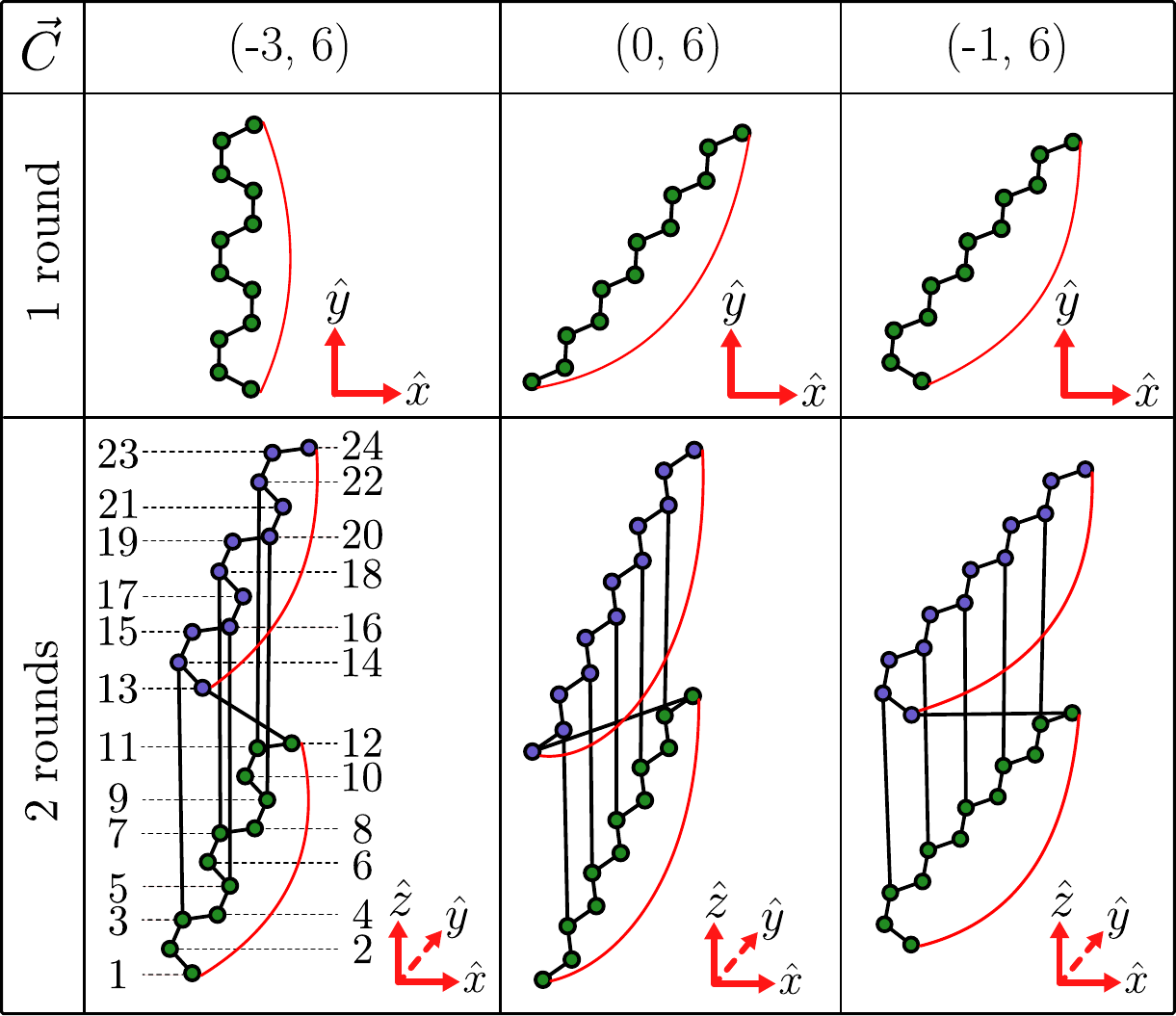}
\caption{
Example of unit cells for one-round and two-round nanoscrolls with chiral vectors $\vec{C}=(-3,6)$, $(0,6)$, and $(-1,6)$.
We define the coordinate system such that the $\hat{x}$-axis is oriented along the axial direction of the nanoscroll, the $\hat{y}$-axis corresponds to the transverse direction of the multilayer ribbon, and the $\hat{z}$-axis is normal to the graphene plane.
Green and purple dots denote lattice sites belonging to different layers. 
The red lines highlight pairs of sites within the same layer in the multilayer representation that become vertically aligned in the rolled-up configuration.
The numbers in the two-round nanoscroll with $\vec{C}=(-3,6)$ correspond to the site labels in Fig.~\ref{fig:scroll_axial_view}.
Each unit cell is repeated infinitely along the axial $\hat{x}$ direction, forming a nanoscroll with translational symmetry.
} 
\label{fig:unit_cells}
\end{figure}

\begin{equation}
\vec{C}= n\vec{a}_1+m\vec{a}_2-\vec{\delta},
\label{Hamiltonian_broken}
\end{equation}
where $\vec{a}_1=(1,0,0)a$ and $\vec{a}_2=(\frac{1}{2}, \frac{\sqrt3}{2}, 0)a$ are the primitive vectors of graphene. 
$\vec{\delta}=(0, \frac{1}{\sqrt{3}}, 0)a$, representing the displacement between A and B sites, is required in the chiral vector definition to satisfy the ABC-stacking condition in our nanoscroll model, as the overlapping atoms in adjacent layers belong to different sublattices.
Here, $a$ is the lattice constant, which we set to $1$ for convenience. 
For simplicity, we denote the chiral vector of a carbon nanoscroll by~$(n, m)$. 
Here $n$ specifies the magnitude of the shift, while $m$ determines the width.
All unshifted structures, such as the one shown in Fig.~\ref{fig:scrolls3d}(a), satisfy $n=-\frac{m}{2}$.

\begin{figure}[tb]
\centering
\includegraphics[width=\columnwidth]{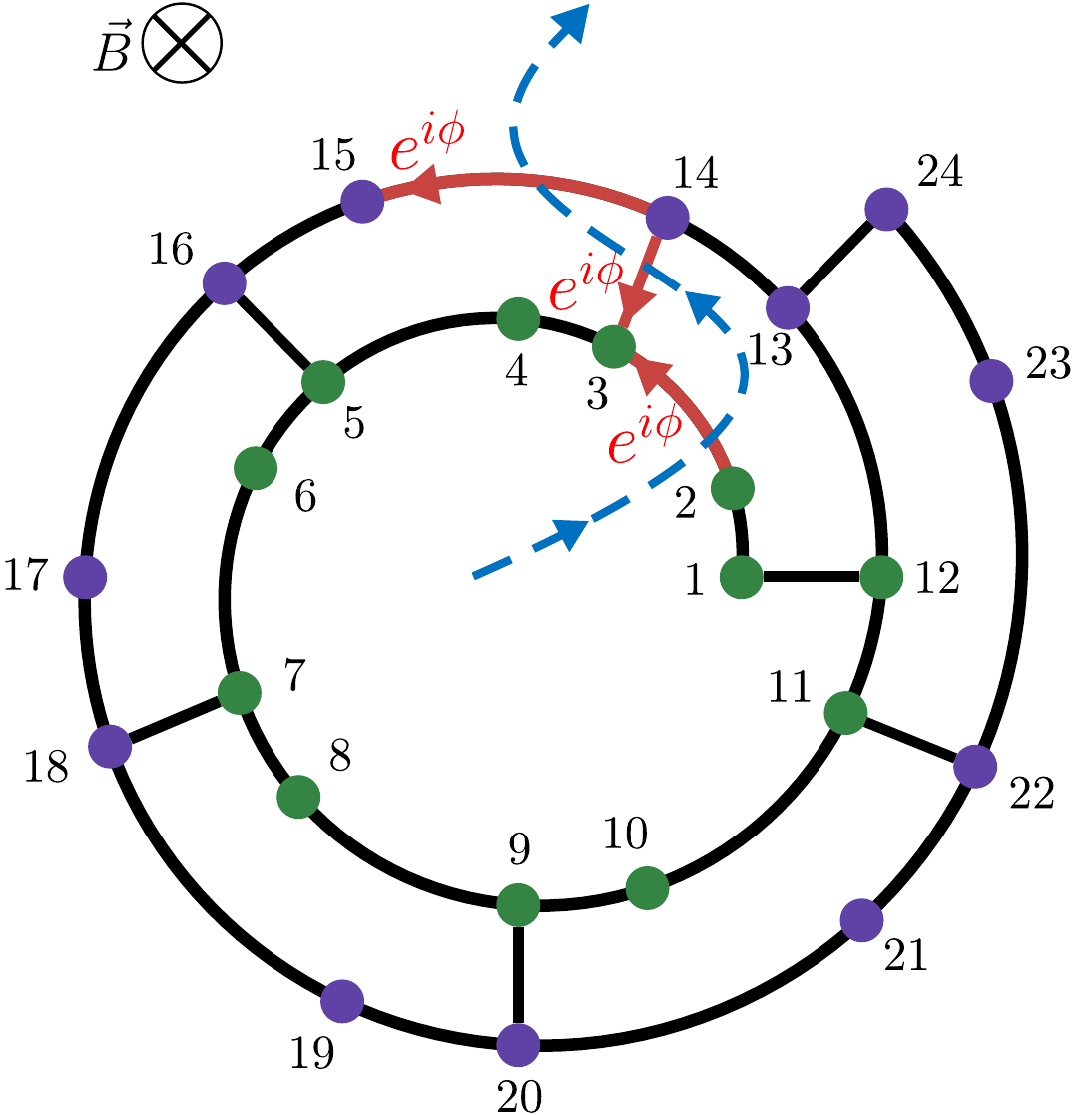}
\caption{
The schematic diagram illustrates the construction of the model and the corresponding Hamiltonian. 
The red and black lines represent the hopping in the structure, while the blue dashed line represents a continuous path extending from the interior of the scroll to infinity. 
The red hopping terms that cross this line acquire the magnetic flux $\phi$.
}
\label{fig:scroll_axial_view}
\end{figure}

One can also roll a single-layer graphene into a multi-round scroll.
Based on the definition of the chiral vector $\vec{C}$, Fig.~\ref{fig:unit_cells} illustrates various unit cells for one- and two-round nanoscrolls of large radius within the multilayer representation. All configurations exhibit translational symmetry along the $\hat{x}$-direction, corresponding to the axial direction of the nanoscroll.
Here we consider the two-round nanoscroll with $\vec{C}=(-3,6)$ in Fig.~\ref{fig:unit_cells} as an example. 
On the flattened multilayer representation, sites connected by $\vec{C}$ within the same layer (e.g., sites 1 and 12, as well as 13 and 24) are mapped onto vertically aligned positions in the rolled-up structure, as illustrated in Fig.~\ref{fig:scroll_axial_view}.
We also note that all two-round nanoscrolls characterized by $\vec{C}=(n,6)$ share the same cross-sectional geometry shown in Fig.~\ref{fig:scroll_axial_view}.

\begin{figure*}[t]
\centering
\includegraphics[width=\textwidth]{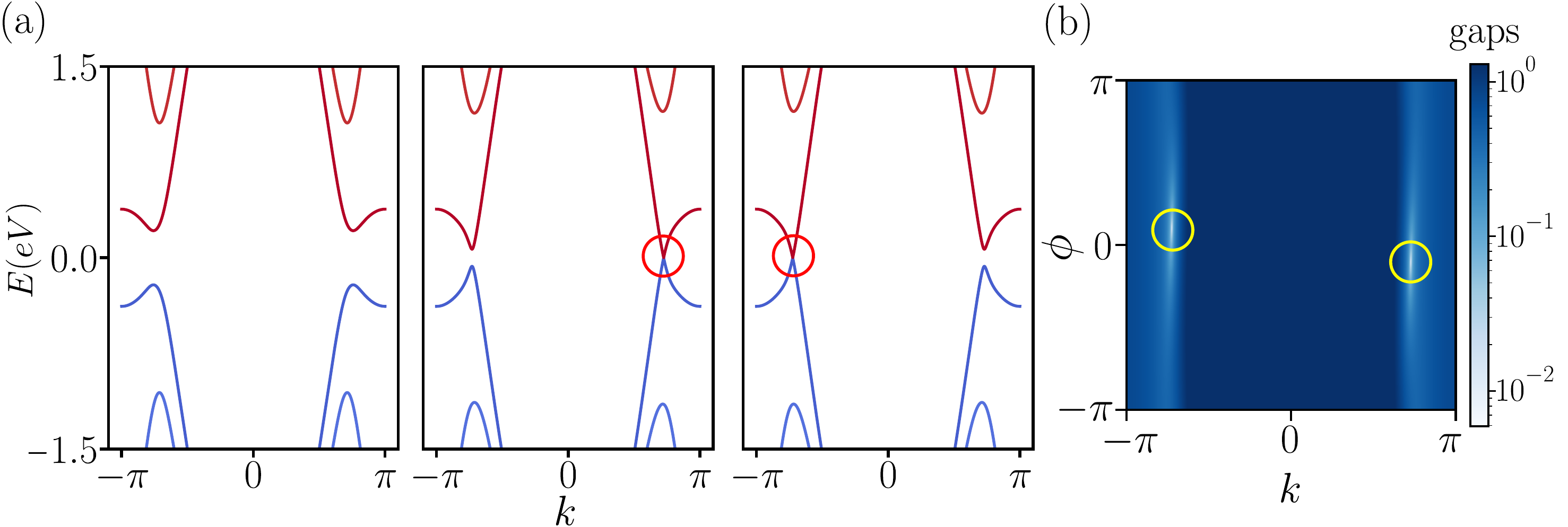} 
\caption{
(a) The band structures of the one-round carbon nanoscroll with chiral vector $\vec{C}=(-2, 12)$ under various magnetic fluxes $\phi$. 
The subfigures, presented from left to right, correspond to $\phi=\pi$, $-0.3188$, and $0.3188$, respectively. 
Unlike unshifted structures, the Dirac cones no longer appear at $\phi = \pi$.
(b) The band gap of the system at the Fermi level $E=0$ is plotted as a function of the momentum $k$ and the magnetic flux $\phi$. 
The yellow circles highlight the two Dirac cones, located approximately at $(k, \phi) \approx (-2.2765, 0.3188)$ and $(2.2765, -0.3188)$.
}
\label{fig:bands_1round}
\end{figure*}

Since the hoppings only connect sites belonging to opposite sublattices (A and B), the system possesses sublattice symmetry, such that the tight-binding Hamiltonian takes a block-off-diagonal form:

\begin{equation}
\label{Hamiltonian}
H(k,\phi)=
\begin{bmatrix}
0 & h(k,\phi)  \\
h^*(k,\phi) & 0
\end{bmatrix},
\end{equation}
where $k$ is the dimensionless momentum along the axial direction and $\phi$ denotes the dimensionless magnetic flux threaded through the nanoscroll.

To illustrate the tight-binding construction, we consider a two-round nanoscroll with $\vec{C}=(-1,6)$ as an example. 
We introduce the basis $\{A_1, A_2, \dots, A_{12}, B_1, B_2, \dots, B_{12}\}$, where $A$ and $B$ denote the sublattices. 
Here, $A_n$ refers to the $2n$-th site, and $B_n$ refers to the $(2n-1)$-th site, respectively, counted from the inside out. 
This labeling follows the site numbering shown in Fig.~\ref{fig:scroll_axial_view}. 
The off-diagonal block $h(k,\phi)$ then reads

\begin{widetext}
\begin{equation}
\label{h}
h(k,\phi) =
\begin{bmatrix}
t_1 & t_2p^* & 0 & 0 & 0 & 0 & 0 & 0 & 0 & 0 & 0 & 0\\
0 & t_1^* & t_2 &0 & 0 & 0 & 0 & 0 & 0 & 0 & 0 & 0\\
0 & 0 & t_1^* & t_2 & 0 & 0 & 0 & 0 & 0 & 0 & 0 & 0\\
0 & 0 & 0 & t_1^* & t_2 & 0 & 0 & 0 & 0 & 0 & 0 & 0 \\
0 & 0 & 0 & 0 & t_1^* & t_2 & 0 & 0 & 0 & 0 & 0 & 0\\
r_1 & 0 & 0 & 0 & 0 & t_1^* & t_2 & 0 & 0 & 0 & 0 & 0\\
0 & r_1p^* & 0 & 0 & 0 & 0 & t_1 & t_2p^* & 0 & 0 & 0 & 0\\
0 & 0 & r_1 & 0 & 0 & 0 & 0 & t_1^* & t_2 & 0 & 0 & 0\\
0 & 0 & 0 & r_1 & 0 & 0 & 0 & 0 & t_1^* & t_2 & 0 & 0\\
0 & 0 & 0 & 0 & r_1 & 0 & 0 & 0 & 0 & t_1^* & t_2 & 0\\
0 & 0 & 0 & 0 & 0 & r_1 & 0 & 0 & 0 & 0 & t_1^* & t_2\\
0 & 0 & 0 & 0 & 0 & 0 & r_1  & 0 & 0 & 0 & 0 & t_1^*
\end{bmatrix},
\end{equation}
\end{widetext}
where $t_2=r_0$, $t_1=r_0(1+e^{ik})$, $t_1^*=r_0(1+e^{-ik})$, $p=e^{i\phi}$ and $p^*=e^{-i\phi}$.  
Here, $\phi = 2\pi\, \Phi/\Phi_0$ is the axial magnetic flux in units of the magnetic flux quantum $\Phi_0$, added by Peierls substitution~\cite{PhysRevLett.108.225303}. 
Since we are working within the large-radius approximation, the magnetic flux contribution associated with the finite wall thickness of the nanoscrolls is negligible.
Due to the spiral geometry of nanoscrolls, we deliberately choose a gauge in which all hopping terms crossing a line extending from the interior of the scroll to infinity acquire the magnetic flux $\phi$, as illustrated by the red hoppings and the blue dashed line in Fig.~\ref{fig:scroll_axial_view}. 
This construction guarantees that only closed loops encircling the origin accumulate the magnetic flux.

\section{One-round scrolls} 
\label{sec:1round}

We begin our analysis with the band structure of a one-round nanoscroll characterized by the chiral vector $\vec{C}=(-2,12)$.
This means that the corresponding unit cell contains 24 sites, and the two edges are shifted with respect to each other by two times the primitive vector $\vec{a}_1$~(see the top panels of Fig.~\ref{fig:unit_cells} for the cases with zero and one shift).
Fig.~\ref{fig:bands_1round}(a) shows the band structures for certain values of $\phi$. 
Dirac cones are observed at $\phi \approx 0.3188$ and $\phi \approx -0.3188$, instead of at $\phi=\pi$, as in Ref.~\cite{Zhong2022-rt}. 
Since the Hamiltonian exhibits $2\pi$ periodicity in $\phi$, it is natural to introduce a fictitious Brillouin zone spanned by $(k,\phi)$. 
Fig.~\ref{fig:bands_1round}(b) shows the energy gap at the Fermi level over the FBZ. 
In this representation, Dirac cones appear as minimum regions~(white) enclosed by the yellow circles. 
The number of Dirac cones agrees with Ref.~\cite{Lee2025-gm}, although their positions in $\phi$ are shifted.

\begin{figure}[t]
\centering
\includegraphics[width=\columnwidth]{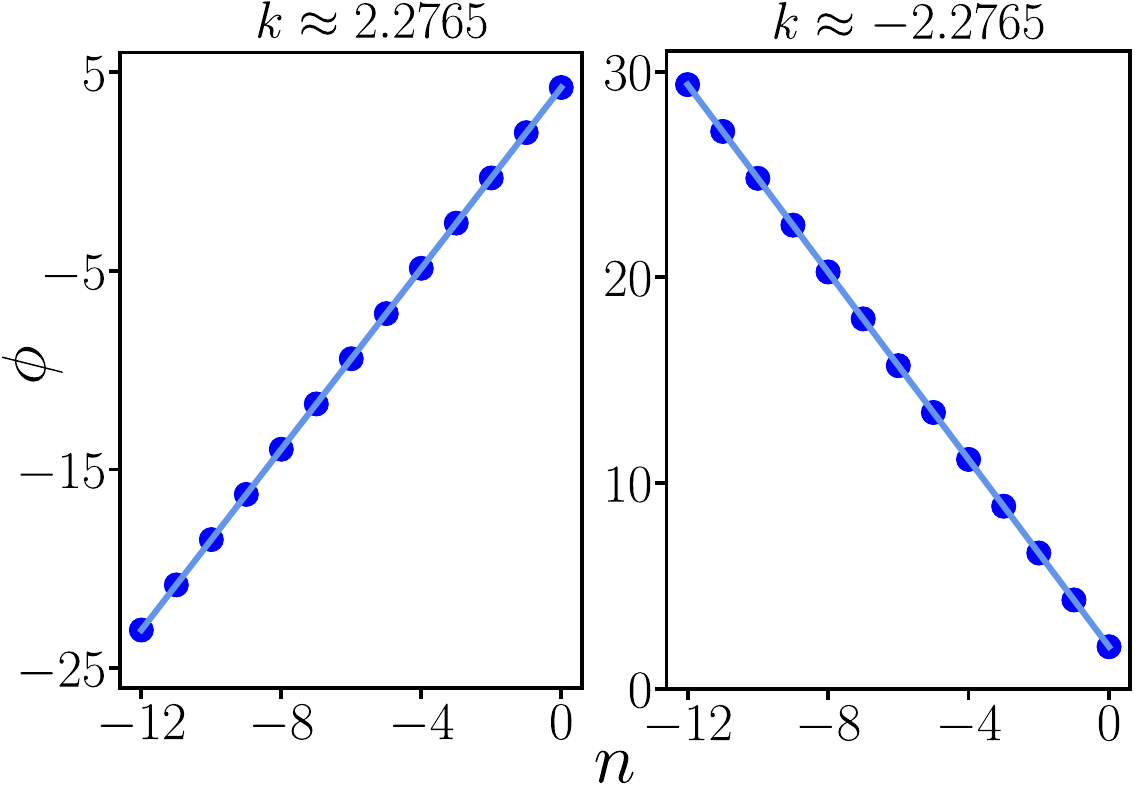} 
\caption{
For one-round carbon nanoscrolls with $m=12$, the Dirac cones are observed at both positive (left) and negative (right) momenta. 
The magnetic flux, $\phi$, at which these Dirac cones appear is observed to exhibit a linear dependence on the chiral index $n$.
}
\label{fig:dirac_pos_1round}
\end{figure}

Varying the chiral vector $(n, m)$ by changing $n$ at fixed $m$, we find that the Dirac cone remains at the same momentum $k$ while the corresponding magnetic flux $\phi$ depends linearly on $n$.
The numerical results for $m=12$ are shown in Fig.~\ref{fig:dirac_pos_1round}, demonstrating this linear dependence.
Note that the Hamiltonian is $2\pi$-periodic in both $k$ and~$\phi$. Consequently, the physical properties remain invariant under the transformation $\phi \rightarrow \phi + 2N\pi$, where $N$ is an integer.
The two subfigures in Fig.~\ref{fig:dirac_pos_1round} exhibit linear fits with opposite slopes, whose magnitude is equal to~$2.2765$, corresponding precisely to the momentum of the Dirac cones.
Extending the analysis to arbitrary $m$, we find that the slope equals the momentum at which the Dirac cone appears.
The relation of the $\phi$, $k$ and $n$ can therefore be written as

\begin{align}
\phi=nk+C,
\label{f(k, phi)}
\end{align}

\noindent
where $C=\pi + m \arccos\left( \frac{1}{2} \left( \frac{-r_1}{r_0} \right)^{\frac{1}{m}} \right)+(2N\pi)$ as derived in App.~\ref{analytical solution for Dirac cones in one-round carbon nanoscrolls}.
This behavior arises because, for fixed $m$, the Hamiltonians of one-round carbon nanoscrolls with different $n$ are related by a flux-shift transformation, 
\begin{equation}
    H_{(n, m)} \big( k, \phi \big)=H_{(n', m)} \big( k, \phi + (n'-n) k \big).
\end{equation}

\begin{figure}[t]
\centering
\includegraphics[width=\columnwidth]{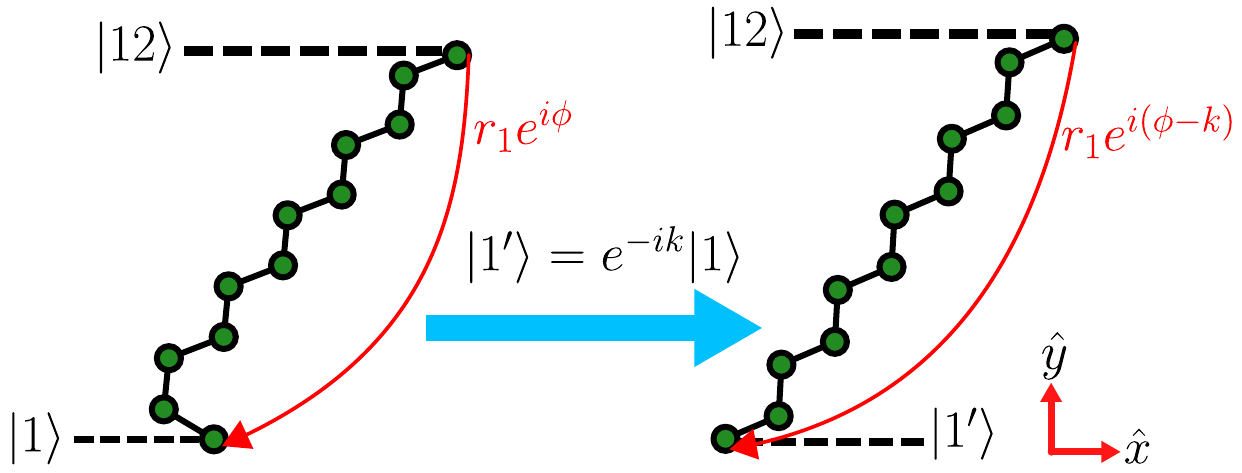} 
\caption{
A one-round chiral carbon nanoscroll with $\vec{C}=(-1, 6)$ can be transformed into one with $\vec{C}=(0, 6)$ by substituting $\phi$ with $\phi-k$ in the phase of the red hopping. This replacement makes the Hamiltonian of the former system identical to that of the latter. Only the boundary hopping terms $\langle 1 | H | 2m \rangle$ and $\langle 2m | H | 1 \rangle$ are modified when transforming the Hamiltonian between different one-round carbon nanoscrolls sharing the same chiral index $m$.
}
\label{fig:different_shift}
\end{figure}

Fig.~\ref{fig:different_shift} illustrates this transformation between the one-round carbon nanoscrolls with $\vec{C}=(-1, 6)$ and $\vec{C}=(0, 6)$. 
Under the conditions described above, the coordinates~$(k, \phi)$ of the Dirac cones in the FBZ for one-round carbon nanoscrolls are derived analytically, as shown in App.~\ref{analytical solution for Dirac cones in one-round carbon nanoscrolls}.

\begin{figure}[t]
\centering
\includegraphics[width=\columnwidth]{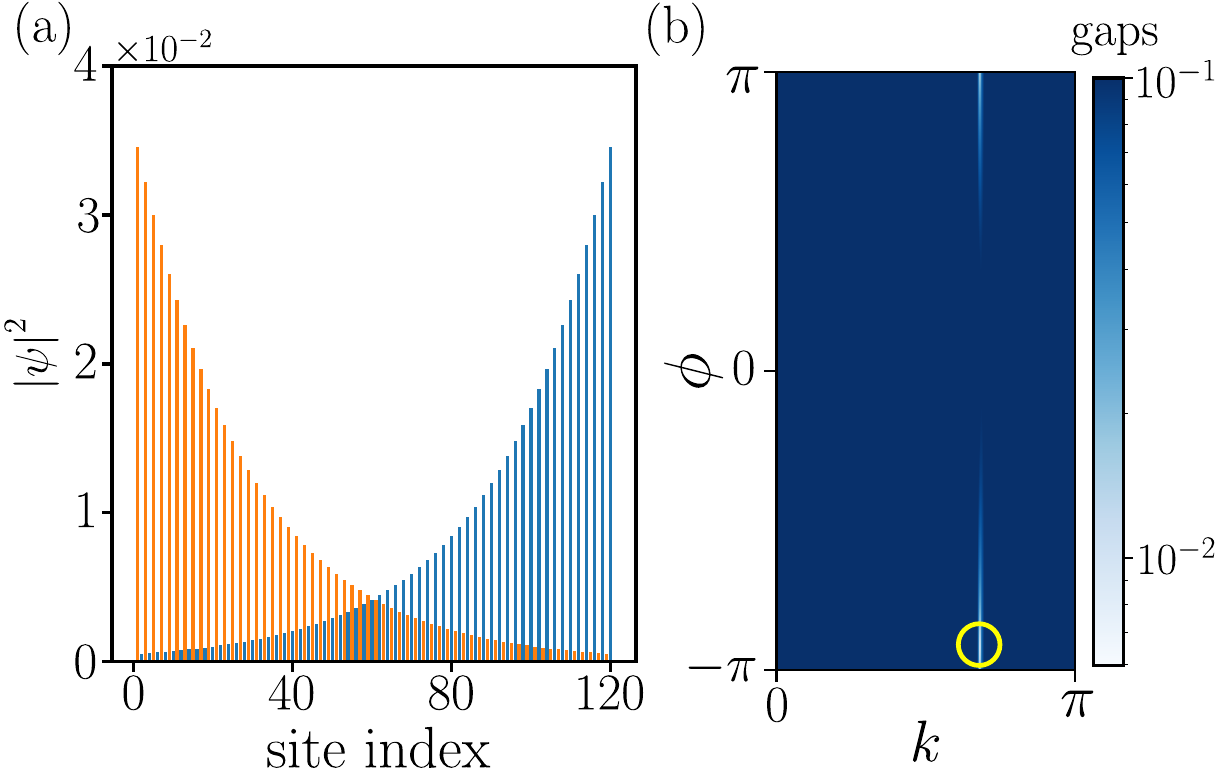}
\caption{
(a) The probability density distribution, $\left| \psi \right|^2$, of the Dirac cone shown in (b), for a one-round carbon nanoscroll with $\vec{C}=(-24,60)$.
The horizontal axis of (a) represents the site index, as shown in Fig.~\ref{fig:scroll_axial_view}. 
The two sublattices, A and B, are distinguished by orange and blue colors, respectively.
}
\label{fig:dirac_wf_1round}
\end{figure}

The Dirac cone in this system retains its bulk state character, consistent with its counterpart in carbon nanotubes. 
The probability density distributions, $\left| \psi \right|^2$, of the Dirac cone at positive momentum are illustrated in Fig.~\ref{fig:dirac_wf_1round} for a one-round carbon nanoscroll with $\vec{C}=(-24,60)$. 
Here, $\left| \psi \right|^2=\frac{1}{2}( \left| \psi_1 \right|^2 + \left| \psi_2 \right|^2)$, where~$\left| \psi_1 \right|^2$ and $\left| \psi_2 \right|^2$ correspond to two degenerate states at Dirac cones.
Notably, the probability density $\left| \psi \right|^2$ is distributed across the entire structure. The distributions on both A and B sublattices exhibit a decay from the boundaries toward the center.

\section{Multi-round scrolls}
\label{sec:multi_round}

\begin{figure*}
\centering
\includegraphics[width=\textwidth]{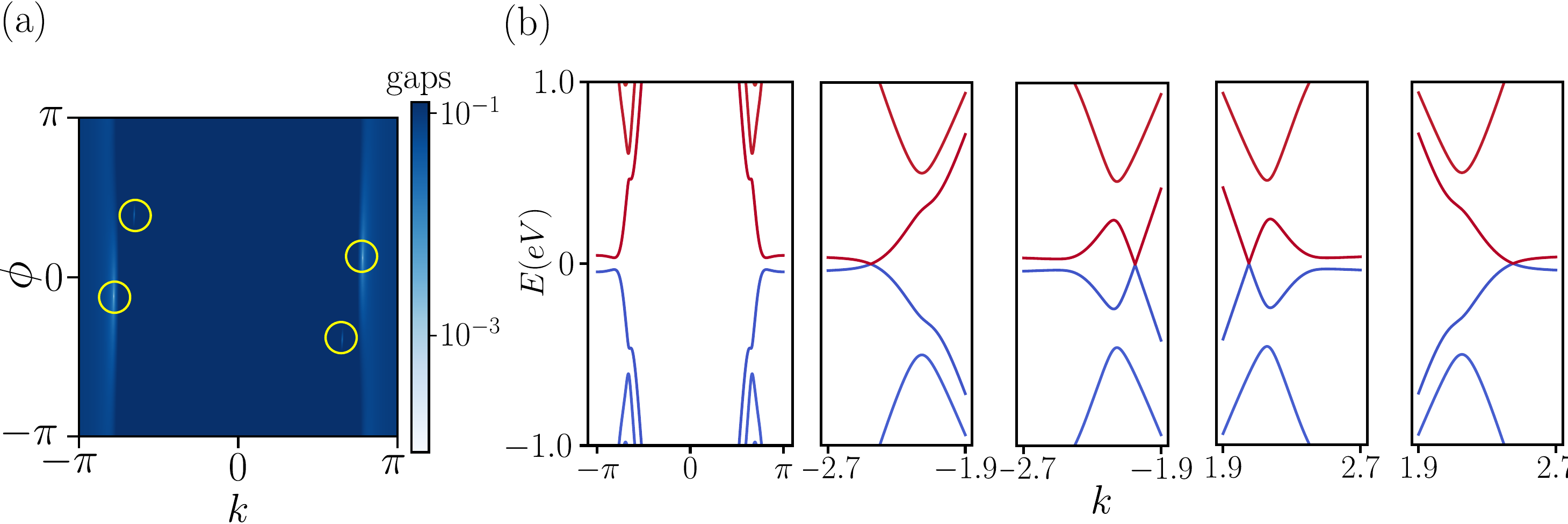} 
\caption{
(a) The energy gaps for the two-round nanoscroll with $\vec{C}=(-2,12)$ presented in the FBZ. 
(b) The band structures under varying magnetic fluxes.
The subfigures, presented from left to right, correspond to $\phi=\pi$, $-0.3775$, $1.222$, $-1.222$, and $0.3775$, respectively. 
} 
\label{fig:bands_2round}
\end{figure*}

We now turn to a two-round scroll, using the chiral vector $\vec{C}=(-2,12)$.
As illustrated in Fig.~\ref{fig:bands_2round}, four Dirac cones appear in the FBZ, consistent with the number found for the unshifted structure in Ref.~\cite{Lee2025-gm}. 
However, similar to the one-round case, introducing a shift in the structure displaces the Dirac cones away from $\phi=\pi$.

\begin{figure*}
\centering
\includegraphics[width=\textwidth]{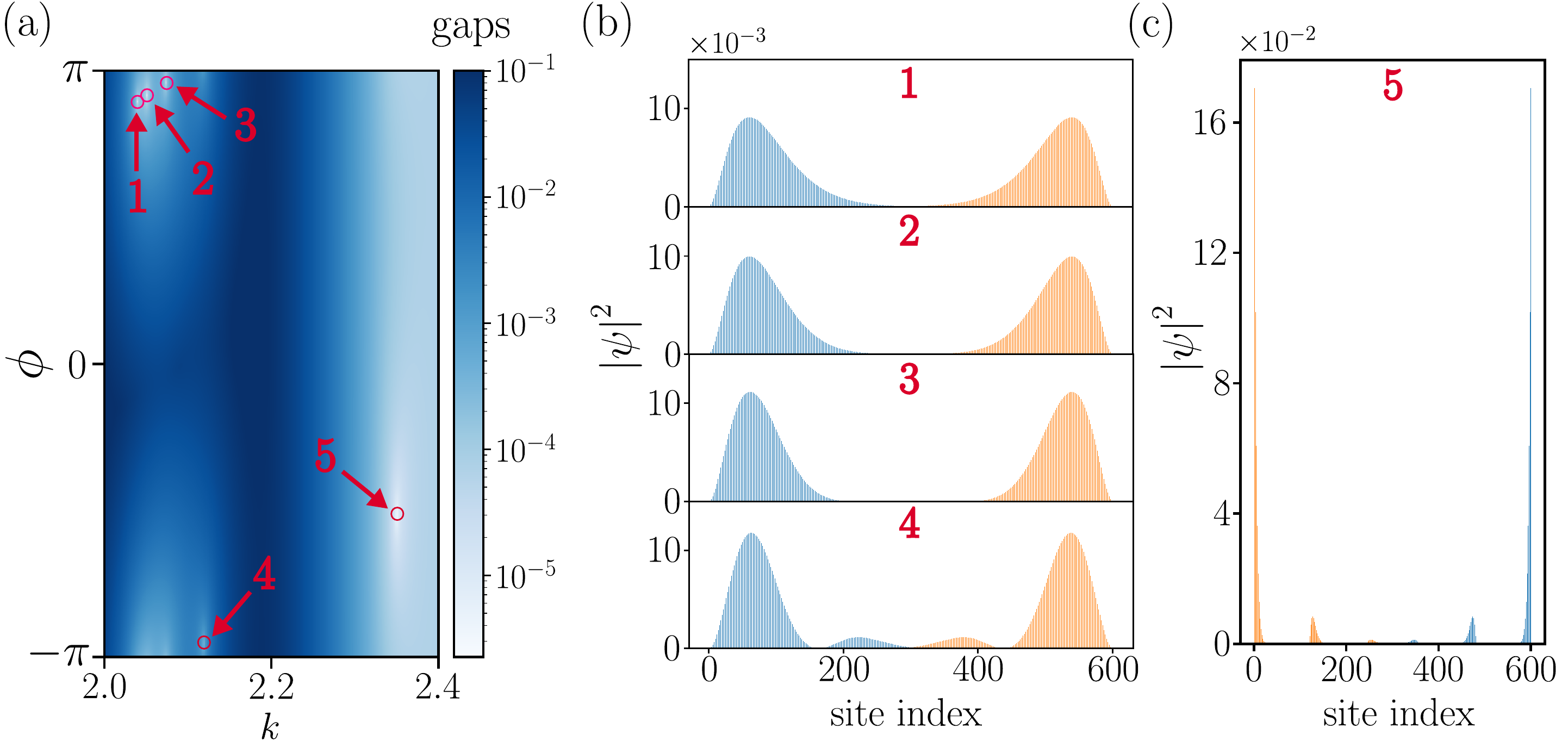} 
\caption{
(a) The energy gaps for the five-round carbon nanoscroll with $\vec{C}=(-24,60)$ in the FBZ. 
The Dirac cones are sequentially labeled by indices 1 through 5. 
Their approximate $(k, \phi)$ coordinates are: (2.0389, 2.8077), (2.0504, 2.8770), (2.0728, 3.0120), (2.1173, -3.0230), and (2.3484, -1.6179), respectively. 
The probability density distributions, $\left| \psi \right|^2$, of the Dirac cones are subsequently shown in (b) for cones 1-4 and (c) for cone 5.
} 
\label{fig:dirac_wf_5round}
\end{figure*}

We find that Dirac cones appearing at different momenta have different real-space probability densities.
To clearly illustrate this, we turn to a five-round carbon nanoscroll characterized by $\vec{C}=(-24,60)$. 
Fig.~\ref{fig:dirac_wf_5round} presents the probability density distribution, $\left| \psi \right|^2$, for the Dirac cones within this structure. 
It is found that the Dirac cone states at larger momentum become more localized at the boundaries. 
However, due to interlayer coupling, the wave function extends into adjacent rounds, leading to additional smaller peaks in the probability density.

\subsection{Symmetry of the carbon nanoscroll}
\label{subsec:sym_list}
 
The Hamiltonian for the carbon nanoscroll, when treated as an effective two-dimensional system in the FBZ that depends on two ``momenta,'' $k$ and $\phi$, exhibits multiple symmetries. 
Specifically, it obeys sublattice symmetry,~$\Gamma$:
\begin{equation}
    \Gamma^\dagger H(k,\phi) \Gamma =- H(k,\phi). \label{eq:sublattice}
\end{equation}
Furthermore, it obeys the spinless time-reversal symmetry, expressed as
\begin{equation}
    H(k,\phi)=H^*(-k,-\phi). \label{eq:trs}
\end{equation}
Moreover, the unshifted structure possesses spinless mirror symmetry along both the $x$ and $y$ directions. The mirror symmetry with respect to the $x$ direction flips the sign of $k$:
\begin{equation}
    M_x^\dagger H(k,\phi) M_x= H(-k,\phi), \label{eq:mx}
\end{equation}
where $M_x$ represents the mirror reflections across the $yz$ plane. This symmetry ensures that $E(k) = E(-k)$ in unshifted carbon nanoscrolls. Similarly, mirror symmetry along the $y$ direction flips the sign of $\phi$:
\begin{equation}
    M_y^\dagger H(k,\phi) M_y= H(k,-\phi), \label{eq:my}
\end{equation}
where $M_y$ denotes the mirror reflection across the $xz$ plane. 
Finally, by combining Eqs.~\eqref{eq:mx} and \eqref{eq:my}, it follows that the Hamiltonian also obeys inversion $I$ symmetry, which simultaneously reverses the signs of both $k$ and $\phi$:
\begin{equation}
    I^\dagger H(k,\phi) I= H(-k,-\phi), \label{eq:inversion}
\end{equation}
Notably, this symmetry is preserved in the shifted structure as well as the unshifted one. Eq.~\eqref{eq:inversion} further explains $E(k, \phi)=E(-k, -\phi)$ that one can observe in the colormap of gaps within the FBZ, such as Fig.~\ref{fig:bands_1round}(b) and Fig.~\ref{fig:bands_2round}(a).

\subsection{Mirror symmetry protection of Dirac cones}
\label{subsec:mir_sym}
\begin{figure}[t]
\centering
\includegraphics[width=\columnwidth]{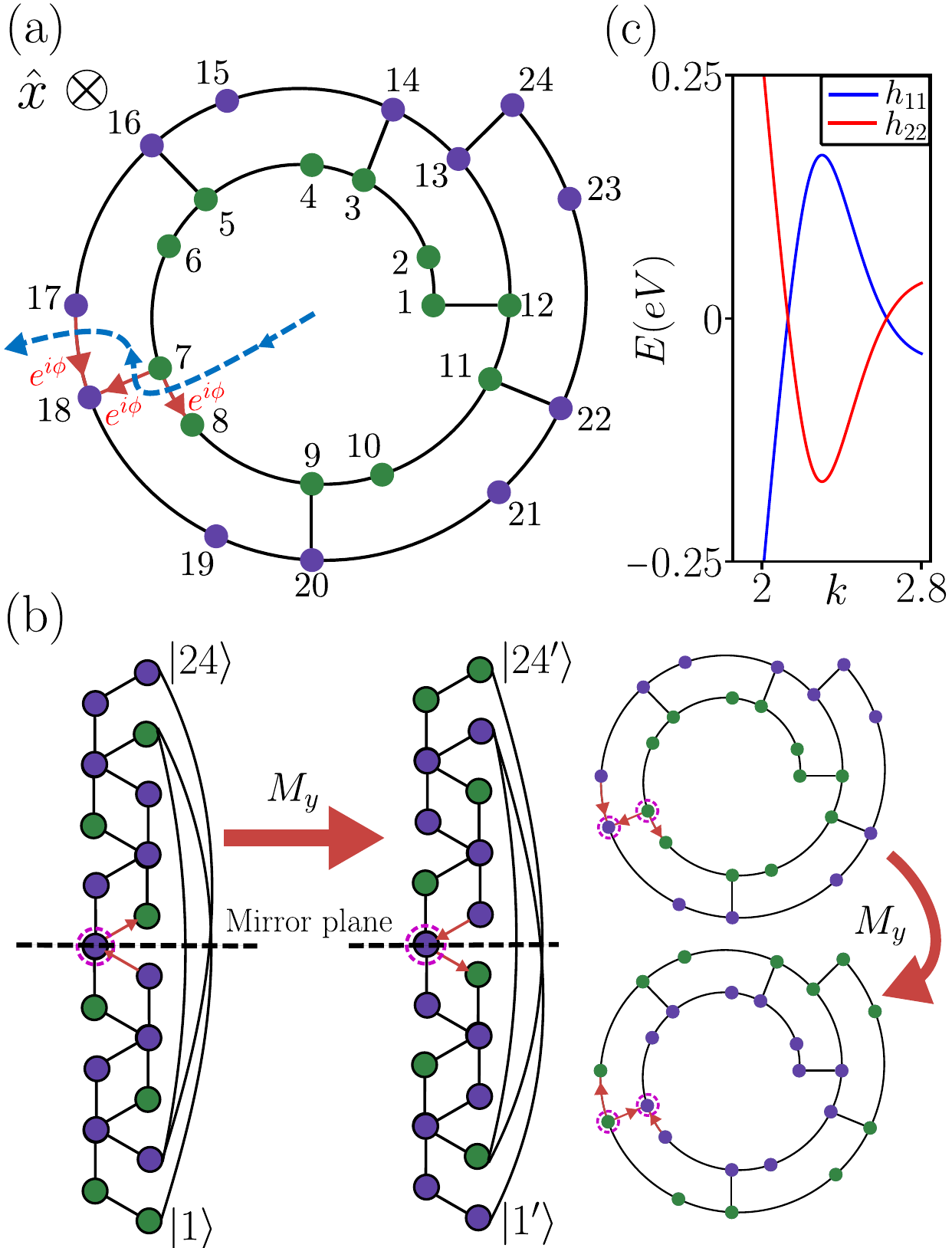}
\caption{
(a) Axial view of the tight-binding model corresponding to a two-round nanoscroll with chiral vector $\vec{C} = (-3, 6)$, including the specific gauge for the axial magnetic flux (red arrows represent the hoppings with the positive phase factor $e^{i\phi}$). 
(b) Left: The ``flattened'' view of the same tight-binding model of the two-round nanoscroll with $\vec{C}=(-3,6)$ in Fig.~\ref{fig:unit_cells}, illustrating the $M_y$ mirror operation and the mirror plane.
Right: The axial view of the two lattices model, before and after applying $M_y$. 
The transformation of the Hamiltonian by $M_y$ is equivalent to redefining the site index starting from the purple site and substituting $\phi$ with $-\phi$.
Note that the redefinition process corresponds to inverting the stacking sequence of the rounds, assuming that $| 1 \rangle$ (or $| 1' \rangle$) and $| 24 \rangle$  (or $| 24' \rangle$) denote the innermost and the outermost sites of the nanoscroll, respectively.
(c) The energy dispersions of $h_{11}$ and $h_{22}$ in Eq.~\eqref{H_my(diagonal)}. 
Each Dirac cone is formed by the intersection of two bands with opposite mirror eigenvalues, demonstrating that they are protected by the $M_y$ symmetry.
}
\label{fig:mirror}
\end{figure}

The mirror symmetry $M_y$ explains why all Dirac cones are aligned at $\phi = \pi$ in unshifted carbon nanoscrolls.
To illustrate this, we take the two-round carbon nanoscroll with $\vec{C} = (-3, 6)$ as an example, and adopt a gauge where the phase factor $e^{i\phi}$ is added only to the hoppings between sites ``7–8,'' ``7–18,'' and ``17–18,'' as shown in Fig.~\ref{fig:mirror}(a).
Fig.~\ref{fig:mirror}(b) illustrates that the $M_y$ operation, represented as a reflection across the mirror plane, is equivalent to redefining the site indices and reversing the sign of the magnetic flux $\phi$ simultaneously.
The matrix $M_y$ is anti-diagonal for this choice of site ordering, and further, all its non-zero entries are $1$ under the specific gauge discussed above. 
One can find a basis that diagonalizes $M_y$ into a block-diagonal form $M_y'$:

\begin{align}
M_y'=
\begin{bmatrix}
\mathbb{I} & 0\\
0 & -\mathbb{I}
\end{bmatrix},
\label{M_y(diagonal)}
\end{align}
where $\mathbb{I}$ is the identity matrix.
Owing to Eq.~\eqref{eq:my} and the relation $H(k,\phi =\pi)=H(k,\phi= -\pi)$, $M_y$ commutes with $H(k,\phi= \pi)$. 
Therefore, the same basis will also block-diagonalize $H(k,\phi= \pi)$:
\begin{align}
H'(k,\pi)=\begin{bmatrix}
h_{11}(k,\pi) & 0\\
0 & h_{22}(k,\pi)
\end{bmatrix},
\label{H_my(diagonal)}
\end{align}
where the two blocks correspond to states with different mirror eigenvalues, $+1$ and $-1$.
The energy dispersions of $h_{11}$ and $h_{22}$ are shown in Fig.~\ref{fig:mirror}(c).
The Dirac cones in the figure are formed by the intersection of one band from $h_{11}$ and one from $h_{22}$, demonstrating that the Dirac cones are protected by the $M_y$ symmetry.

\subsection{\texorpdfstring{$IT$}{IT} symmetry protection of Dirac cones}
\label{subsec:c2t_sym}
\begin{figure}[t]
\centering
\includegraphics[width=\columnwidth]{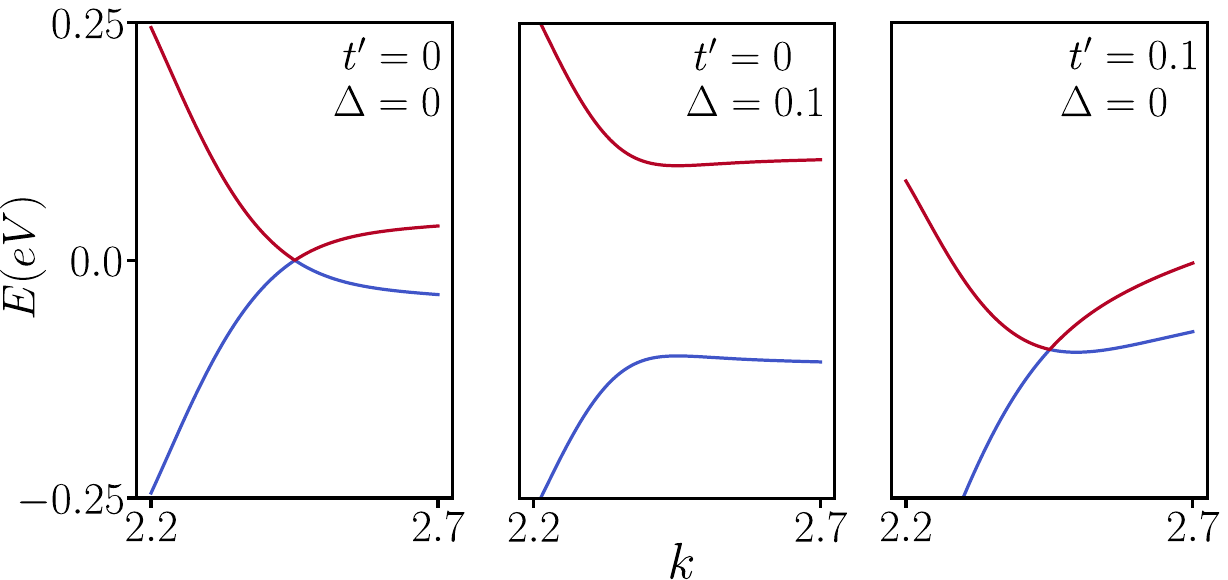}
\caption{
The band structures around the Fermi level of the two-round carbon nanoscroll with $\vec{C}=(-2,12)$. 
The Dirac point persists upon adding a nonzero $t'$, but is gapped out with $\Delta \neq 0$.
}
\label{fig:break_chiral}
\end{figure}

In shifted carbon nanoscrolls, the mirror symmetry is broken; however, the Dirac cones are preserved.
To understand the robustness of the Dirac cones in the absence of mirror symmetry, we examine the remaining symmetries of the system, namely sublattice, spinless time-reversal, and inversion symmetries, as defined in Eqs.~\eqref{eq:sublattice},~\eqref{eq:trs}, and~\eqref{eq:inversion}.
To demonstrate that sublattice symmetry is not essential for the protection of the Dirac cones, we introduce perturbations that break sublattice symmetry, such as an AB potential difference~($V_a = -V_b = \Delta \neq 0$ ) or next-nearest-neighbor intralayer hopping ($t'$).
The results are shown in Fig.~\ref{fig:break_chiral}. 
We find that an AB sublattice potential difference, which breaks inversion symmetry, introduces a mass term and thus gaps the Dirac cones. 
In contrast, next-nearest-neighbor hopping $t'$, which preserves inversion symmetry, does not gap out the Dirac cones.
This behavior can be understood in terms of the combined inversion–time-reversal symmetry (IT), under which the Hamiltonian satisfies
\begin{equation}
    (IT)^{\dagger} H(k,\phi) (IT) = H(k,\phi),\label{c2t}
\end{equation}
where $T={\cal K}$ is the spinless time-reversal operator, equal to complex conjugation. 
This symmetry protects the Dirac cones even in the absence of sublattice symmetry.
To make this symmetry more transparent, we introduce a transformed Hamiltonian
\begin{equation}\label{eq:H_real_gauge}
    \tilde{H}(k,\phi)=\sqrt{I}^{\dagger}H(k,\phi)\sqrt{I},
\end{equation}
where $\sqrt{I}$ is a unitary and symmetric matrix satisfying $I=(\sqrt{I})^2$. Under this basis transformation, $\tilde{H}$ is a real Hamiltonian that follows
\begin{equation}
    \tilde{H}(k,\phi)= \tilde{H}^*(k,\phi).
\end{equation}
The detailed derivation is provided in App.~\ref{prove H' is real}.

As discussed in Refs.~\cite{Bouhon2019, Bouhon2020, PhysRevX.9.021013}, two-dimensional spinless systems obeying the combination of spatial inversion and time-reversal symmetry can host topologically protected Dirac cones in their Brillouin zone, which are characterized by a nonzero winding number even in the absence of sublattice symmetry.
We show that this protection is applicable also to the band crossings of the nanoscrolls, owing to the two-dimensional nature of the FBZ.
Working in the real gauge of Eq.~\eqref{eq:H_real_gauge}, the presence of a nonzero winding number, and hence the protected nature of the Dirac cones, can be seen by the formation of so-called ``Dirac strings'' that emanate from these band crossings.
The strings are one-dimensional regions in $(k, \phi)$ space across which there is a jump in the real-valued wave function characterizing the Dirac bands, highlighting the impossibility of constructing a basis which is both real and continuous around the band crossing.

\begin{figure}[t]
\centering
\includegraphics[width=\columnwidth]{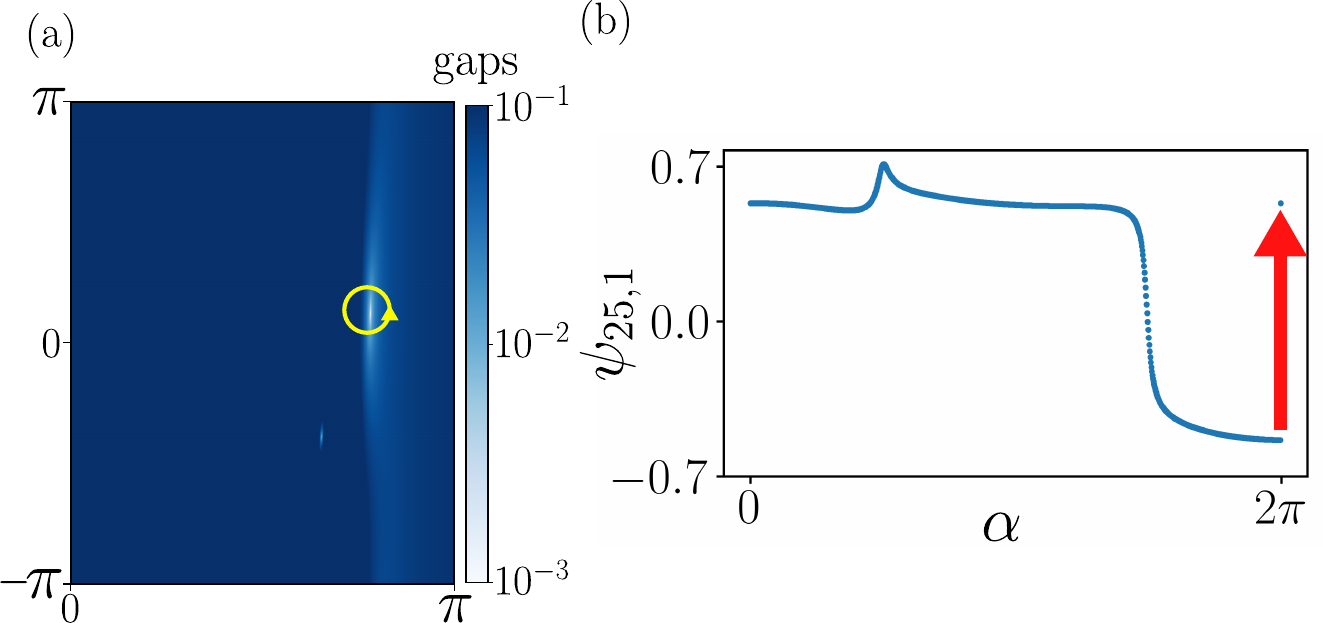}
\caption{
(a) Dirac cones appear in two-round carbon nanoscrolls with $\vec{C}=(-2,12)$. We consider a circular path around one of the Dirac cones, marked with a yellow circle and parameterized by an angle $\alpha \in [0, 2\pi)$. (b) The first component of the eigenvector of the middle band (band index 25) along the path is plotted with $\alpha$. 
The red arrow highlights the discontinuity that originates from crossing a Dirac string.
}
\label{fig:continuous_gauge}
\end{figure}

We show an example of such a Dirac string in Fig.~\ref{fig:continuous_gauge}(a), for the case of a two-round nanoscroll with $\vec{C}=(-2,12)$.
We diagonalize the Hamiltonian along a circular path surrounding one of the band crossings of this system (shown in yellow), parameterized by an angle $\alpha \in [0, 2\pi)$. Then, we implement a smoothing procedure~(see App.~\ref{app:gauge_fixing}) in order to obtain a continuously varying, real wave function~\cite{PhysRevB.85.115415}.
After one full loop, $\alpha \to 2\pi$, we observe that the wave function has the opposite sign compared to that at $\alpha=0$. 
This discontinuity is the Dirac string that emanates from the band crossing, which shows that the latter is protected by $IT$ symmetry, independently of sublattice symmetry.
This additional topological protection explains why in Fig.~\ref{fig:break_chiral} the Dirac cone persists upon adding next-nearest-neighbor hopping (which preserves $IT$), but becomes gapped upon adding a staggered sublattice potential (which breaks both sublattice symmetry and $IT$).
Note that this phenomenon is absent for loops that do not encircle Dirac cones; in that case, it is possible to find wave functions that are both real and continuous (see App.~\ref{app:gauge_fixing}).

\setlength{\belowcaptionskip}{-10pt}

\section{Conclusion} 
\label{sec:conclusions}

In this work, we investigate the Dirac cones in chiral carbon nanoscrolls within a simple, spinless tight-binding model in the large-radius approximation. 
We show that the roll-up direction, characterized by the chiral vector, breaks mirror symmetry and shifts the Dirac cones to magnetic flux values $\phi$ away from half a flux quantum, in contrast to previous reports for unshifted structures~\cite{Lee2025-gm}.
We use the axial momentum and the magnetic flux threading the nanoscroll to define the coordinates $(k, \phi)$ of an effectively two-dimensional fictitious Brillouin zone (FBZ), which provides a natural framework for understanding both the location and the topological protection of the Dirac cones. 

For one-round nanoscrolls, we derive an analytical expression linking the chiral vector to the momentum and the magnetic flux at which the Dirac cones appear.
In the case of multi-round scrolls, we observe that the number of Dirac cones in the FBZ is twice the number of rounds, consistent with previous results~\cite{Lee2025-gm}, regardless of the roll-up direction.
Further, the real-space probability density of their wave functions varies across distinct Dirac cones: states at smaller momenta are more bulk-like, whereas states at larger momenta are more localized at the edges of the nanoscroll.
In contrast, the Dirac cone in the one-round carbon nanoscroll is a bulk state, which can be viewed as being inherited from graphene.

To understand the robustness of the gapless points, we analyzed the symmetries of the nanoscrolls.
In unshifted structures, the mirror symmetry along the $y$-direction not only protects the Dirac cones but also constrains them to occur at a magnetic flux of $\phi = \pi$, corresponding to half a flux quantum.
Moreover, in shifted structures, even when the sublattice symmetry is broken by introducing next-nearest-neighbor hopping, the Dirac cones persist. 
We demonstrated that their protection in these cases stems from the underlying $IT$ symmetry, a concept verified through eigenvector analysis.

This work provides a comprehensive model of chiral carbon nanoscrolls, detailing the coordinates and state character of the Dirac cones. 
Most importantly, we found that the Dirac cones in our carbon nanoscroll models are generically protected by $IT$ symmetry, rather than just sublattice symmetry, hinting at the possibility that they may be observed in experiments. 
This work contributes to our understanding of carbon nanoscrolls and establishes a basis for future research into their potential applications. 
Moreover, the methodology and model proposed are applicable to other two-dimensional materials.

\section*{Acknowledgements} 
\label{sec:acknowledgements}

We gratefully thank Ching-Hao Chang for the valuable discussions.
T.-C. H. and J.-S. Y. acknowledge support from the National Science and Technology Council (NSTC), Taiwan, under Grant No. NSTC 113-2112-M-003-015 -, and from “Higher Education Sprout Project” of National Taiwan Normal University and the Ministry of Education (MOE), Taiwan.  J.-S. Y. and H.-C. H. acknowledge the support from the National Center for Theoretical Sciences (NCTS) in Taiwan. T.-C. H. acknowledges support from the New Generation Talent Cultivation Scholarship provided by the Department of Physics, National Taiwan Normal University. ICF was funded by the German Science Foundation (Deutsche Forschungsgemeinschaft, DFG) in the framework of the Cluster of Excellence on Complexity and Topology in Quantum Matter ct.qmat (EXC 2147, Project No. 390858490).

\appendix
\section{Analytical solution for Dirac cones in one-round carbon nanoscrolls}
\label{analytical solution for Dirac cones in one-round carbon nanoscrolls} 

To determine the constant $C$ in Eq.~\eqref{f(k, phi)}, we consider the special case of an unshifted structure, characterized by the chiral vector $\vec{C} = (-m/2, m)$. For such configurations, it is established that the Dirac cones are located at $\phi = \pi$. By adopting a gauge where the magnetic flux is assigned to the $r_1$ hopping terms, the determinant of the Hamiltonian is derived as
\begin{equation}
\det(H(k, \phi=\pi)) = \left[ (2r_0 \cos(k/2))^m + r_0^{m-1}r_1 \right]^2.
\label{eq:det_analytical}
\end{equation}
Since the determinant of a Hamiltonian is equal to the product of all its eigenenergies, and the Dirac cones correspond to zero-energy states due to sublattice symmetry, the following condition must be satisfied:
\begin{equation}
\det(H_{(-\frac{m}{2},m)}(k=k', \phi=\pi)) = 0,
\label{eq:det_zero}
\end{equation}
where $k'$ denotes the momentum of the Dirac cones.
Solving Eq.~\eqref{eq:det_zero}, we obtain one of the solutions for $k'$,
\begin{equation}
k' = 2 \arccos\left( \frac{1}{2} \left( \frac{-r_1}{r_0} \right)^{1/m} \right).
\label{eq:k_solution}
\end{equation}
As Eq.~\eqref{eq:det_analytical} is an even function of $k$ with $2\pi$ periodicity, the general solutions are given by $\pm(k' + 2n\pi)$ for $n \in \mathbb{Z}$. By utilizing Eq.~\eqref{eq:k_solution} and the condition that Dirac cones appear at $\phi=\pi$ for the unshifted structure, $C$ in Eq.~\eqref{f(k, phi)} is determined to be
\begin{equation}
C = \pi + m \arccos\left( \frac{1}{2} \left( \frac{-r_1}{r_0} \right)^{\frac{1}{m}} \right).
\label{eq:Cm_val}
\end{equation}
Consequently, substituting Eq.~\eqref{eq:Cm_val} back to Eq.~\eqref{f(k, phi)}, we find that for a one-round carbon nanoscroll, the Dirac cones appear at
\begin{equation}
\begin{cases}
k' = 2 \arccos\left( \frac{1}{2} \left( \frac{-r_1}{r_0} \right)^{\frac{1}{m}} \right) \\
\phi' = \pi + (m + 2n) \arccos\left( \frac{1}{2} \left( \frac{-r_1}{r_0} \right)^{\frac{1}{m}} \right).
\end{cases}
\label{eq:final_result}
\end{equation}

\section{Proof that \texorpdfstring{$\tilde{H}$}{H} is real}
\label{prove H' is real}

To prove that $\tilde{H} = \sqrt{I}^{\dagger} H \sqrt{I}$ is a real-valued matrix, we divide the derivation into four stages: (1) establishing that $I$ is a symmetric matrix; (2) applying Takagi decomposition ~\cite{192483} to factorize $I = VV^{T}$, where $V$ is a unitary matrix; (3) demonstrating that the transformed Hamiltonian $\tilde{H} = V^{\dagger} H V$ is real; and (4) identifying $V$ as the square root of $I$.

\noindent
(1) Establishing that $I$ is a symmetric matrix\\

In our spinless model, the $IT$ symmetry is represented by an anti-unitary operator that squares to plus one: $(IT)^2 = \mathbb{I}$. This identity implies that the matrix condition $I I^* = \mathbb{I}$. Since $I$ is also a unitary matrix, it follows that
\begin{equation}
I^* = I^{\dagger} \implies I = I^{T},
\end{equation}
which confirms that $I$ is a symmetric matrix.\\

\noindent
(2) Applying Takagi decomposition~\cite{192483} to factorize $I = VV^{T}$, where $V$ is a unitary matrix\\

According to the Takagi decomposition, any symmetric matrix can be decomposed as $VDV^{T}$, where $V$ is a unitary matrix and $D$ is a non-negative diagonal matrix. Therefore, we can express the symmetric matrix $I$ as\\
\begin{equation}
    I=VDV^{T}.\\ 
\end{equation}
Using the unitary property, it follows that\\
\begin{equation}
    VDV^{T}V^{*}DV^{\dagger}=VD^{2}V^{\dagger}=\mathbb{I}.\\
\end{equation}
This identity implies that $D^{2}=\mathbb{I}$.
Given that $D$ is a non-negative diagonal matrix, we must have $D=\mathbb{I}$. Consequently, $I$ can be factorized as
\begin{equation}
    I=VV^T \label{takagi c2}.\\
\end{equation}

\noindent
(3) Demonstrating that the transformed Hamiltonian $\tilde{H} = V^{\dagger} H V$ is real\\

We cast the $IT$ symmetry condition from Eq.~\eqref{c2t} into the form:
\begin{equation}
    I^{\dagger} H I = H^*
\end{equation}
By substituting Eq.~\eqref{takagi c2} into the equation above, we obtain 
\begin{equation}
    V^{*}V^{\dagger}HVV^{T}=H^{*}.\\
\end{equation}
To isolate the transformed Hamiltonian, we multiply by $V^T$ from the left and $V^*$ from the right. This leads to
\begin{equation}
    V^{\dagger}HV=V^{T}H^{*}V^{*}=(V^{\dagger}HV)^{*}\\
\end{equation}
By defining the transformed Hamiltonian as $\tilde{H} = V^{\dagger} H V$, it follows that
\begin{equation}
    \tilde{H}=\tilde{H}^{*},\\
\end{equation}
which proves that $\tilde{H}$ is a real matrix.

\noindent
(4) Identifying $V$ as the square root of the $I$ operator.\\

Let $\{u_n\}$ be the set of eigenvectors of $I$, satisfying:
\begin{equation}
I u_n = e^{i\theta_n} u_n.
\end{equation}
Taking the complex conjugate of both sides and utilizing the property that $I$ is a symmetric unitary matrix, we obtain:
\begin{equation}
I^{-1} u_n^* = e^{-i\theta_n} u_n^* \implies I u_n^* = e^{i\theta_n} u_n^*.
\end{equation}
Since $u_n$ and $u_n^*$ share the same eigenvalue, their linear combinations span the same eigenspace. We can therefore construct a real basis, such as $(u_n + u_n^*)/\sqrt{2}$, to form a real unitary matrix $U$, where $U=U^*$. In this basis, $I$ is diagonalized as
\begin{equation}
I = U \Lambda U^\dagger,
\end{equation}
where $\Lambda = \text{diag}(e^{i\theta_1}, \dots, e^{i\theta_N})$. By defining the square root as $\sqrt{I} = U \sqrt{\Lambda} U^\dagger$, it follows that
\begin{equation}
(\sqrt{I})^2 = I \quad \text{and} \quad \sqrt{I} = (\sqrt{I})^T.
\end{equation}
Consequently, the relation $I = \sqrt{I} \sqrt{I}^T$ holds, confirming that $\sqrt{I}$ is a valid representation for the matrix $V$ in the Takagi decomposition.

\section{Gauge smoothing of eigenvectors}
\label{app:gauge_fixing}

\begin{figure}[h]
\centering
\includegraphics[width=\columnwidth]{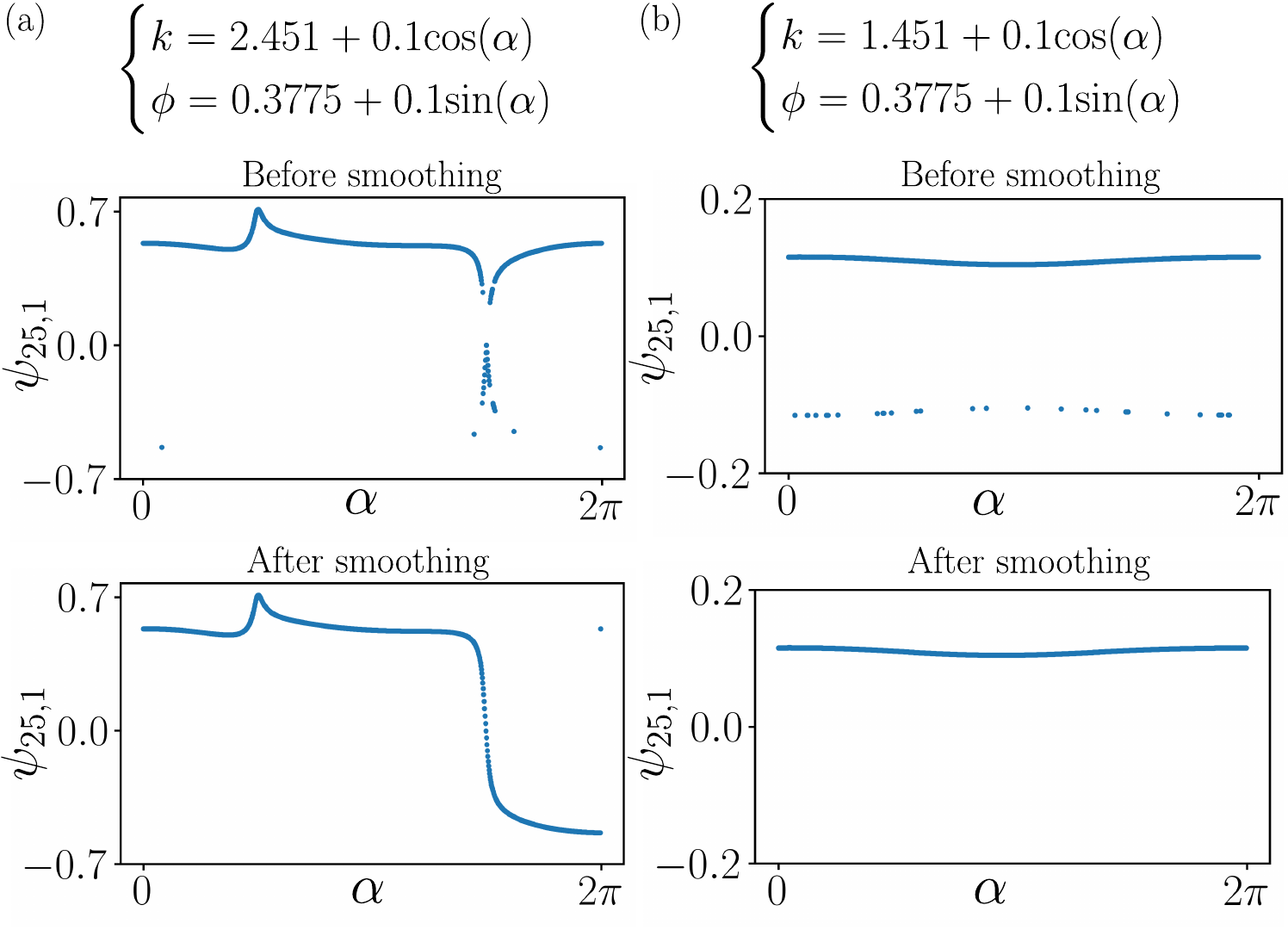}
\caption{
The evolution of the eigenvector components along two distinct paths: one enclosing the Dirac cone (a) and another avoiding it (b), following the same format as Fig.~\ref{fig:continuous_gauge}. The top panels specify the equations for each trajectory, while the middle and bottom panels present the corresponding components before and after the gauge-smoothing procedure, respectively.
}
\label{fig:gauge_smoothing}
\end{figure}

When numerically diagonalizing a Hamiltonian along a path in parameter space (here $k$ and $\phi$), it is not a priori guaranteed that the eigenvectors will change continuously. 
This is due to the fact that eigenvectors are defined only up to a global phase. 
Here, since we work with the Hamiltonian in a real gauge, the eigenvectors are also real, and the only allowed global phases are 0 or $\pi$. 
We obtain a smooth gauge by ensuring that the overlap $\langle\psi(\alpha_i)|\psi(\alpha_{i+1})\rangle$, for adjacent points $\alpha_i$ along the path, is real and positive. To this end, we manually flip the sign of $|\psi(\alpha_{i+1})\rangle$ if the overlap is negative. 
The results before and after smoothing with different paths are demonstrated in Figs.~\ref{fig:gauge_smoothing}(a) and (b).

\newpage

\bibliography{references}

\end{document}